\documentclass[cpc,amsmath,amssymb,showpacs,superscriptaddress,nofootinbib,twocolumn]{revtex4-1}
\usepackage{graphicx}
\usepackage{epstopdf}
\usepackage{dcolumn}
\usepackage{color}
\usepackage{bm}
\usepackage{CJK,CJKnumb,CJKulem}     
\usepackage{overpic}
\usepackage{rotating}
\usepackage{times}
\usepackage{multirow}
\usepackage{float}
\usepackage{mathrsfs}
\usepackage{booktabs}
\usepackage[mathlines]{lineno}

\newcommand{\EE}{e^+e^-}

\newcommand{\LL}{\ell^+\ell^-}

\newcommand{\pipipi}{\pi^+\pi^-\pi^0}
\newcommand{\KK}{K^+K^-}

\newcommand{\y}{\Upsilon(10860)}
\newcommand{\yy}{\Upsilon(11020)}

\newcommand{\B}{\mathcal{B}}

\begin{document}
\hyphenpenalty=10000
\tolerance=2000

\graphicspath{{figure/}}
\DeclareGraphicsExtensions{.eps,.png,.ps}


\title{\quad\\[2.0cm] \boldmath
Observation of $\EE\to \pipipi\chi_{b1,2}(1P)$ and search for
$\EE\to\phi\chi_{b1,2}(1P)$ at $\sqrt{s}=$10.96---11.05 GeV}
\vspace{0.2cm}

\noaffiliation
\affiliation{University of the Basque Country UPV/EHU, 48080 Bilbao}
\affiliation{Beihang University, Beijing 100191}
\affiliation{University of Bonn, 53115 Bonn}
\affiliation{Brookhaven National Laboratory, Upton, New York 11973}
\affiliation{Budker Institute of Nuclear Physics SB RAS, Novosibirsk 630090}
\affiliation{Faculty of Mathematics and Physics, Charles University, 121 16 Prague}
\affiliation{Chonnam National University, Kwangju 660-701}
\affiliation{University of Cincinnati, Cincinnati, Ohio 45221}
\affiliation{Deutsches Elektronen--Synchrotron, 22607 Hamburg}
\affiliation{University of Florida, Gainesville, Florida 32611}
\affiliation{Key Laboratory of Nuclear Physics and Ion-beam Application (MOE) and Institute of Modern Physics, Fudan University, Shanghai 200443}
\affiliation{Justus-Liebig-Universit\"at Gie\ss{}en, 35392 Gie\ss{}en}
\affiliation{II. Physikalisches Institut, Georg-August-Universit\"at G\"ottingen, 37073 G\"ottingen}
\affiliation{SOKENDAI (The Graduate University for Advanced Studies), Hayama 240-0193}
\affiliation{Gyeongsang National University, Chinju 660-701}
\affiliation{Hanyang University, Seoul 133-791}
\affiliation{University of Hawaii, Honolulu, Hawaii 96822}
\affiliation{High Energy Accelerator Research Organization (KEK), Tsukuba 305-0801}
\affiliation{J-PARC Branch, KEK Theory Center, High Energy Accelerator Research Organization (KEK), Tsukuba 305-0801}
\affiliation{Forschungszentrum J\"{u}lich, 52425 J\"{u}lich}
\affiliation{IKERBASQUE, Basque Foundation for Science, 48013 Bilbao}
\affiliation{Indian Institute of Technology Bhubaneswar, Satya Nagar 751007}
\affiliation{Indian Institute of Technology Guwahati, Assam 781039}
\affiliation{Indian Institute of Technology Hyderabad, Telangana 502285}
\affiliation{Indian Institute of Technology Madras, Chennai 600036}
\affiliation{Indiana University, Bloomington, Indiana 47408}
\affiliation{Institute of High Energy Physics, Chinese Academy of Sciences, Beijing 100049}
\affiliation{Institute of High Energy Physics, Vienna 1050}
\affiliation{Institute for High Energy Physics, Protvino 142281}
\affiliation{INFN - Sezione di Napoli, 80126 Napoli}
\affiliation{INFN - Sezione di Torino, 10125 Torino}
\affiliation{Advanced Science Research Center, Japan Atomic Energy Agency, Naka 319-1195}
\affiliation{J. Stefan Institute, 1000 Ljubljana}
\affiliation{Institut f\"ur Experimentelle Teilchenphysik, Karlsruher Institut f\"ur Technologie, 76131 Karlsruhe}
\affiliation{Kennesaw State University, Kennesaw, Georgia 30144}
\affiliation{King Abdulaziz City for Science and Technology, Riyadh 11442}
\affiliation{Department of Physics, Faculty of Science, King Abdulaziz University, Jeddah 21589}
\affiliation{Korea Institute of Science and Technology Information, Daejeon 305-806}
\affiliation{Korea University, Seoul 136-713}
\affiliation{Kyoto University, Kyoto 606-8502}
\affiliation{Kyungpook National University, Daegu 702-701}
\affiliation{LAL, Univ. Paris-Sud, CNRS/IN2P3, Universit\'{e} Paris-Saclay, Orsay}
\affiliation{\'Ecole Polytechnique F\'ed\'erale de Lausanne (EPFL), Lausanne 1015}
\affiliation{P.N. Lebedev Physical Institute of the Russian Academy of Sciences, Moscow 119991}
\affiliation{Faculty of Mathematics and Physics, University of Ljubljana, 1000 Ljubljana}
\affiliation{Ludwig Maximilians University, 80539 Munich}
\affiliation{Luther College, Decorah, Iowa 52101}
\affiliation{University of Malaya, 50603 Kuala Lumpur}
\affiliation{University of Maribor, 2000 Maribor}
\affiliation{Max-Planck-Institut f\"ur Physik, 80805 M\"unchen}
\affiliation{School of Physics, University of Melbourne, Victoria 3010}
\affiliation{University of Mississippi, University, Mississippi 38677}
\affiliation{University of Miyazaki, Miyazaki 889-2192}
\affiliation{Moscow Physical Engineering Institute, Moscow 115409}
\affiliation{Moscow Institute of Physics and Technology, Moscow Region 141700}
\affiliation{Graduate School of Science, Nagoya University, Nagoya 464-8602}
\affiliation{Kobayashi-Maskawa Institute, Nagoya University, Nagoya 464-8602}
\affiliation{Universit\`{a} di Napoli Federico II, 80055 Napoli}
\affiliation{Nara Women's University, Nara 630-8506}
\affiliation{National Central University, Chung-li 32054}
\affiliation{National United University, Miao Li 36003}
\affiliation{Department of Physics, National Taiwan University, Taipei 10617}
\affiliation{H. Niewodniczanski Institute of Nuclear Physics, Krakow 31-342}
\affiliation{Nippon Dental University, Niigata 951-8580}
\affiliation{Niigata University, Niigata 950-2181}
\affiliation{Novosibirsk State University, Novosibirsk 630090}
\affiliation{Osaka City University, Osaka 558-8585}
\affiliation{Pacific Northwest National Laboratory, Richland, Washington 99352}
\affiliation{Panjab University, Chandigarh 160014}
\affiliation{Peking University, Beijing 100871}
\affiliation{University of Pittsburgh, Pittsburgh, Pennsylvania 15260}
\affiliation{Theoretical Research Division, Nishina Center, RIKEN, Saitama 351-0198}
\affiliation{University of Science and Technology of China, Hefei 230026}
\affiliation{Showa Pharmaceutical University, Tokyo 194-8543}
\affiliation{Soongsil University, Seoul 156-743}
\affiliation{Stefan Meyer Institute for Subatomic Physics, Vienna 1090}
\affiliation{Sungkyunkwan University, Suwon 440-746}
\affiliation{School of Physics, University of Sydney, New South Wales 2006}
\affiliation{Department of Physics, Faculty of Science, University of Tabuk, Tabuk 71451}
\affiliation{Tata Institute of Fundamental Research, Mumbai 400005}
\affiliation{Excellence Cluster Universe, Technische Universit\"at M\"unchen, 85748 Garching}
\affiliation{Department of Physics, Technische Universit\"at M\"unchen, 85748 Garching}
\affiliation{Toho University, Funabashi 274-8510}
\affiliation{Department of Physics, Tohoku University, Sendai 980-8578}
\affiliation{Earthquake Research Institute, University of Tokyo, Tokyo 113-0032}
\affiliation{Department of Physics, University of Tokyo, Tokyo 113-0033}
\affiliation{Tokyo Institute of Technology, Tokyo 152-8550}
\affiliation{Tokyo Metropolitan University, Tokyo 192-0397}
\affiliation{Virginia Polytechnic Institute and State University, Blacksburg, Virginia 24061}
\affiliation{Wayne State University, Detroit, Michigan 48202}
\affiliation{Yamagata University, Yamagata 990-8560}
\affiliation{Yonsei University, Seoul 120-749}
  \author{J.~H.~Yin}\affiliation{Institute of High Energy Physics, Chinese Academy of Sciences, Beijing 100049} 
  \author{C.~Z.~Yuan}\affiliation{Institute of High Energy Physics, Chinese Academy of Sciences, Beijing 100049} 
  \author{I.~Adachi}\affiliation{High Energy Accelerator Research Organization (KEK), Tsukuba 305-0801}\affiliation{SOKENDAI (The Graduate University for Advanced Studies), Hayama 240-0193} 
  \author{H.~Aihara}\affiliation{Department of Physics, University of Tokyo, Tokyo 113-0033} 
  \author{S.~Al~Said}\affiliation{Department of Physics, Faculty of Science, University of Tabuk, Tabuk 71451}\affiliation{Department of Physics, Faculty of Science, King Abdulaziz University, Jeddah 21589} 
  \author{D.~M.~Asner}\affiliation{Brookhaven National Laboratory, Upton, New York 11973} 
  \author{V.~Aulchenko}\affiliation{Budker Institute of Nuclear Physics SB RAS, Novosibirsk 630090}\affiliation{Novosibirsk State University, Novosibirsk 630090} 
  \author{T.~Aushev}\affiliation{Moscow Institute of Physics and Technology, Moscow Region 141700} 
  \author{R.~Ayad}\affiliation{Department of Physics, Faculty of Science, University of Tabuk, Tabuk 71451} 
  \author{V.~Babu}\affiliation{Tata Institute of Fundamental Research, Mumbai 400005} 
% ...
\collaboration{The Belle Collaboration}

\begin{abstract}
We report searches for the processes $e^+e^-\to\pi^+\pi^-\pi^0\chi_{bJ}$ and $e^+e^-\to\phi\chi_{bJ}$ (J=1,2) based on data samples collected by the Belle experiment at the KEKB collider.
We report the first observation of the process $e^+e^-\to
(\pi^+\pi^-\pi^0)_{\rm non-\omega}\chi_{b1}$ and first evidence for $e^+e^-\to\omega\chi_{bJ}$ in the vicinity of the $\Upsilon(11020)$ resonance, with
center-of-mass energies from 10.96 to 11.05~GeV.
The significances for $(\pi^+\pi^-\pi^0)_{\rm non-\omega}\chi_{b1}$ and $\omega\chi_{bJ}$ are greater than $5.3\sigma$ and $4.0\sigma$, respectively.
We also investigate the energy dependence of the $e^+e^-\to\pi^+\pi^-\pi^0\chi_{bJ}$ cross section,
but we cannot determine whether the contributions are from the $\Upsilon(10860)$ and $\Upsilon(11020)$ resonances or non-resonant continuum processes.
The signals for $\EE\to \phi\chi_{bJ}$ are not significant, and the upper
limits of the Born cross sections at the 90\% confidence
level are $0.7$ and $1.0$~pb for $\EE\to
\phi\chi_{b1}$ and $\phi\chi_{b2}$, respectively, for
center-of-mass energies from 10.96 to 11.05~GeV.

\end{abstract}

\maketitle

Hadronic transitions among heavy quarkonium states serve as a
key source of information for better understanding the
strong interaction between a quark and antiquark, and thus
quantum chromodynamics (QCD). The heavy quarkonium systems, in which the speed of quarks is sufficiently small, are approximately non-relativistic,
and the hadronic transitions to lower lying
states have long been described using the QCD multipole
expansion~\cite{multpole}. However, the existence of
anomalously large hadronic transition rates from the $\Upsilon(10860)$, as
reported by the Belle experiment~\cite{hadtrans1,hadtrans2,hadtrans3,hadtrans5,hadtrans6,hadtrans7,hadtrans8,pipihb},
challenges the theoretical calculations, as well as the pure
bottomonium nature of the $\Upsilon(10860)$ and $\Upsilon(11020)$~\cite{bound,FourQuark,BBmix}.

\begin{table*}[!htbp]
\caption{The predicted branching fractions of $\yy\to\omega\chi_{bJ}$ and $\phi\chi_{bJ}$~\cite{Theory}, as well as the relative magnitudes, where $\B_{j}\equiv\B(\yy\to\omega(\phi)\chi_{bj}),~R_{ij}\equiv\frac{\B_{i}}{\B_{j}}$.}\label{tab:prediction}
\begin{tabular}{ c c c c c c c }
  \hline
  \hline
  Decay mode            & $\B_{0}$    &    $\B_{1}$   & $\B_{2}$   &   $R_{10}$ &    $R_{20}$  &     $R_{21}$   \\
  \hline
  $\omega\chi_{bJ}$ ~~& ~~$(0.15\sim 2.81)\times10^{-3}$ ~~&~~  $(0.63\sim 11.68)\times10^{-3}$  ~~&~~   $(1.08\sim20.02)\times10^{-3}$  ~~&~~$\approx4.11$~~&~~$\approx7.06$~~&~~$\approx1.72$ \\
  $\phi\chi_{bJ}$   ~~& ~~$(0.68\sim 4.62)\times10^{-6}$ ~~&~~  $(0.50\sim 3.43)\times10^{-6}$   ~~&~~   $(2.22\sim 15.18)\times10^{-6}$ ~~&~~$\approx0.74$~~&~~$\approx3.28$~~&~~$\approx4.43$ \\
  \hline
\end{tabular}
\end{table*}

The processes $\EE\to \omega\chi_{bJ}$ were observed
recently~\cite{hadtrans3} using data samples taken at energies near
the $\Upsilon(10860)$ peak, but the dependence of the $\EE\to \omega\chi_{bJ}$ cross section versus energy was not
measured. Therefore, it is unclear whether this process occurs from the $\Upsilon(10860)$ meson or continuum process.
Nevertheless, the result has been investigated extensively by theorists to
understand the dynamics of these transitions, producing studies of
$S$- and $D$-wave mixing for the observed heavy quark spin-symmetry violation from the comparison of $\omega\chi_{b1}$ and $\omega\chi_{b2}$~\cite{guofk}, a possible
contribution of $\Upsilon(10860)\to \pi Z_b\to \pi\rho\Upsilon(1S)$~\cite{lix},
a molecular component in the $\Upsilon(10860)$ wave function~\cite{lix}, and
hadronic-loop effects~\cite{chendy}.

%

By extending the calculation in Ref.~\cite{chendy} to the $\yy$ case, assuming the hadronic-loop effect is a universal mechanism in the higher bottomonium transitions, the authors of Ref.~\cite{Theory} predict the branching fractions
of $\yy\to\omega\chi_{bJ}$ in addition to $\yy\to\phi\chi_{bJ}$, where $J=0,~1,$ and $2$, as listed in Table~\ref{tab:prediction}.
Relative magnitudes of these branching fractions are also predicted (and listed in Table~\ref{tab:prediction}), which are weakly dependent on the free parameters introduced in the theoretical calculation.
An experimental measurement of these $\omega$ and $\phi$ transitions will give a crucial test on how
well the hadronic-loop effect works in $\yy$ decay, and a test of the similarity between $\yy$ and $\y$.

In this paper, we report the results of a
search for $\omega\chi_{bJ}$ and
$\phi\chi_{bJ}$ using the $\Upsilon(10860)$ and $\Upsilon(11020)$ energy scan data collected
with the Belle detector.
The data that we are using consist of 22 samples of high integrated luminosity (listed in Table I),
and 18 additional samples of about $50~\rm pb^{-1}$ per point taken in 5 MeV steps between 10.96 and 11.05 GeV~\cite{dataset}.
We use $\chi_{bJ}\to \gamma\Upsilon(1S)$,
$\Upsilon(1S)\to \LL~(\ell=e,~\mu)$, $\omega\to \pipipi$ to
reconstruct the $\EE\to \omega\chi_{b1,2}$ signal; for the $\EE\to
\phi\chi_{bJ}$ signal, we reconstruct $\phi$ with its decays to
$\KK$ and check the production of $\chi_{bJ}$ by studying the
$\KK$ recoil mass.

\begin{table}[!htbp]
\caption{Integrated luminosity at different c.m. energy as well as the energy-dependent Born cross sections for $\EE\to
\pipipi\chi_{bJ}$ with statistical uncertainty only. A 11.9\% common systematic uncertainty is not included.}\label{tab:XS}
\begin{tabular}{ c c c c  }
  \hline
  \hline   $~~E_{\rm c.m.}$ (GeV)~~ &~~$\mathcal{L}(\rm~fb^{-1})$~~&~~~$\sigma^{\rm Born}(\pipipi\chi_{bJ})$ ~(pb)~~\\
  \hline
  10.7711        &  0.955 &  $ -1.44^{+2.62}_{-1.74} $\\
  10.8203        &  1.164 &  $  2.72^{+2.07}_{-1.43} $\\
  10.8497        &  0.989 &  $  2.70^{+2.19}_{-1.41} $\\
  10.8589        &  0.989 &  $  0.64^{+1.51}_{-0.75} $\\
  10.8633        & 47.648 &  $  0.82^{+0.10}_{-0.10} $\\
  10.8667        & 45.553 &  $  0.68^{+0.10}_{-0.10} $\\
  10.8686        & 22.938 &  $  0.89^{+0.16}_{-0.16} $\\
  10.8695        &  0.978 &  $  1.23^{+1.96}_{-1.21} $\\
  10.8785        &  0.978 &  $  1.90^{+1.90}_{-1.17} $\\
  10.8836        &  1.230 &  $  1.37^{+1.56}_{-1.01} $\\
  10.8889        &  0.989 &  $  1.20^{+1.63}_{-0.93} $\\
  10.8985        &  0.983 &  $  1.14^{+1.55}_{-0.88} $\\
  10.9011        &  0.873 &  $ -1.25^{+1.82}_{-1.06} $\\
  10.9077        &  0.980 &  $  0.51^{+1.50}_{-0.87} $\\
  10.9275        &  0.667 &  $  2.12^{+2.11}_{-1.30} $\\
  10.9575        &  0.851 &  $  0.70^{+1.67}_{-0.83} $\\
  10.9775        &  0.999 &  $  2.84^{+1.96}_{-1.32} $\\
  10.9919        &  0.986 &  $  1.10^{+1.50}_{-0.87} $\\
  11.0068        &  0.976 &  $  3.05^{+1.86}_{-1.28} $\\
  11.0164        &  0.771 &  $  3.47^{+2.11}_{-1.46} $\\
  11.0175        &  0.849 &  $  0.00^{+0.95}_{-0.32} $\\
  11.0220        &  0.982 &  $  0.84^{+1.49}_{-0.98} $\\
  \hline
\end{tabular}
\end{table}

The Belle detector, located at the KEKB asymmetric-energy $\EE$ collider~\cite{kekb} is
described in Ref.~\cite{belle}.
The {\sc evtgen}~\cite{evtgen} generator, as well as a {\sc geant3}~\cite{geant}-based detector simulation, is used
to produce simulated events using Monte Carlo (MC) methods.
The nominal parameters of the states in the decay chains are quoted from Ref.~\cite{PDG}.
To take the initial-state radiation (ISR) into consideration, the radiator function from Ref~\cite{ISRfun} are introduced in {\sc
evtgen}.
A generic MC sample at the $\y$ peak including all possible decays is used to study the
possible background channels and investigate the background shape.

For charged tracks, the impact parameters perpendicular to and
along the beam direction with respect to the interaction point are
required to be less than 1.0 and 3.5~cm, respectively. The
transverse momentum is restricted to be higher than 0.1~GeV/$c$.
A particle identification (PID) hypothesis~\cite{PID} $\mathcal{L}(X)$ for each charged track is formed
from different detector subsystems for particle
$X\in{e, \mu, \pi, K, p}$. Tracks with a likelihood ratio
$\mathcal{R}(K)=\mathcal{L}(K)/(\mathcal{L}(K)+\mathcal{L}(\pi))<0.4$ are identified as pions
while those with $\mathcal{R}(K)>0.6$ are identified as kaons.
Similarly, we define the likelihood ratios ${\cal R}(e)$ and ${\cal R}(\mu)$ for identification of electrons and muons, respectively, with ${\cal R}(e) > 0.01$ and ${\cal R}(\mu) > 0.1$. A neutral cluster in the electromagnetic
calorimeter is reconstructed as a photon if it does not match the
extrapolated position of any charged track and its energy is
greater than 30~MeV.

To select $\EE\to \pipipi\chi_{bJ}$ candidates, we require that
there be exactly four tracks, of which two are
positively identified as pions and the other two as leptons. At least three
photons are required in the event, and a $\pi^0$ list is
created with the invariant mass of the photon pairs satisfying
$M(\gamma\gamma) \in[0.12,~0.15]$~GeV/$c^2$, which covers nearly $\pm3\sigma$ around the $\pi^0$ peak.
To improve the track momentum and photon energy resolutions, and to suppress the
background, a five-constraint (5C) kinematic fit is performed for
the $\gamma\pipipi\LL$ candidates enforcing energy and momentum conservation and constraining
the invariant mass of $\pi^0$ candidates.
The four-momenta of the final-state particles after the 5C kinematic fit are kept for further analysis.
The $\chi^2_{\rm 5C}/\rm ndf$ is required to be less than
20, where $\chi^2_{\rm 5C}$ is the
resulting $\chi^2$ of the kinematic fit, and $\rm ndf=5$ is the number of degrees of freedom.
If there are multiple $\pi^0$ candidates surviving the kinematic fit in an event, the one with the smallest
$\chi^2_{\rm 5C}$ is kept.
The lepton pair is taken as an $\Upsilon(1S)$ candidate if its invariant mass is in the region [9.42, 9.60]~GeV/$c^2$.

\begin{figure}[htbp]
\centering
    \includegraphics[width=0.2391\textwidth]{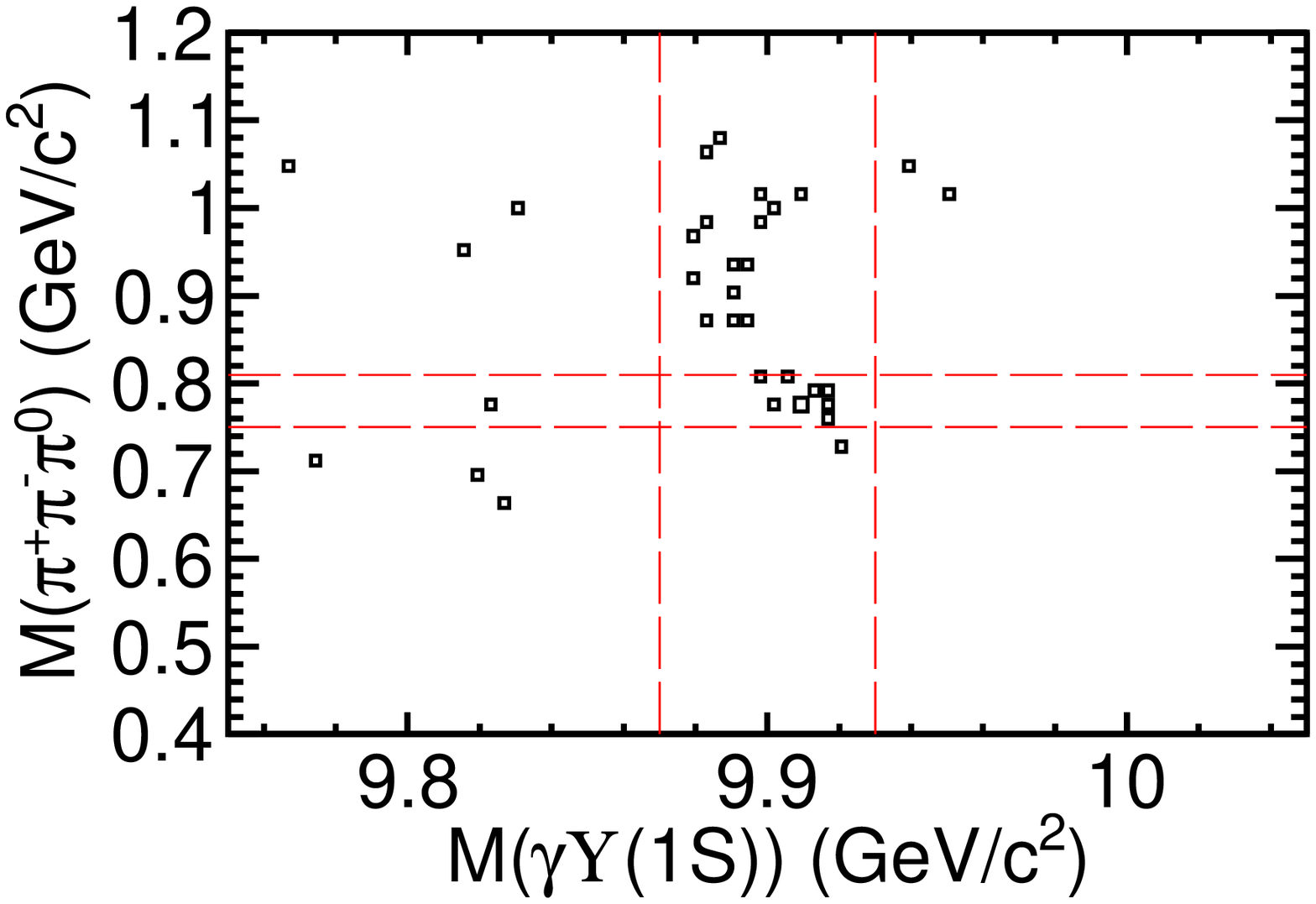}
    \includegraphics[width=0.2391\textwidth]{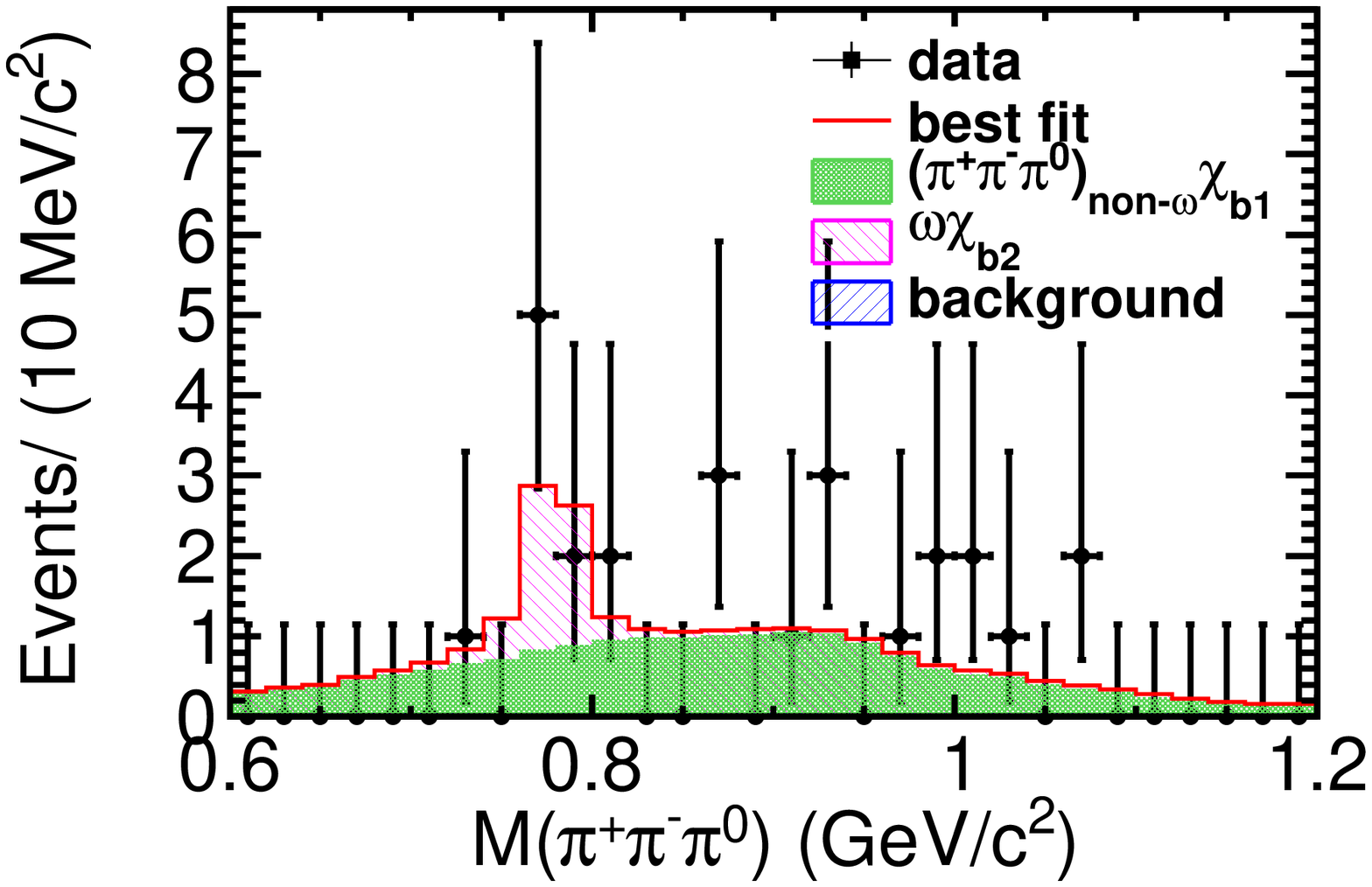}
    \includegraphics[width=0.2391\textwidth]{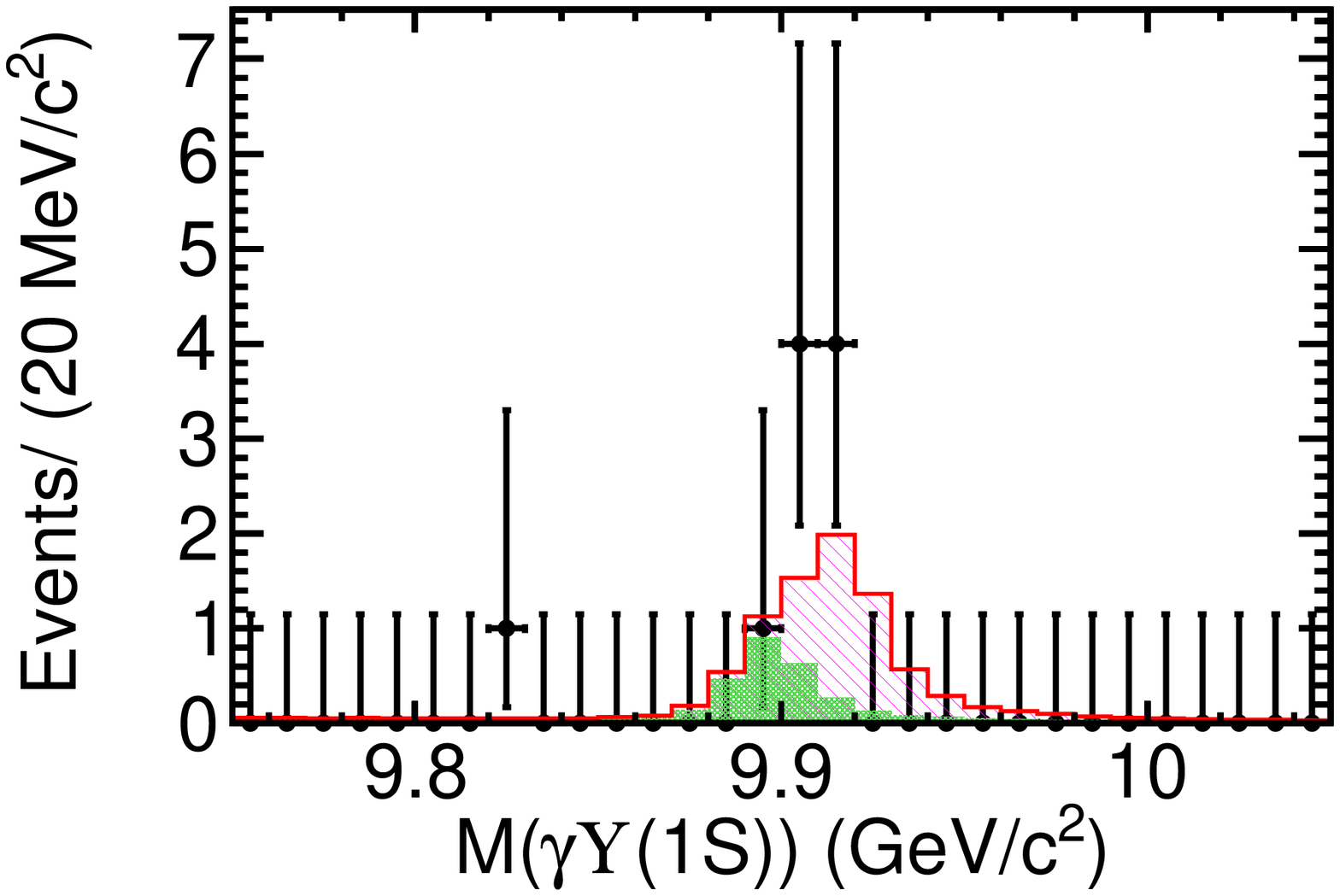}
    \includegraphics[width=0.2391\textwidth]{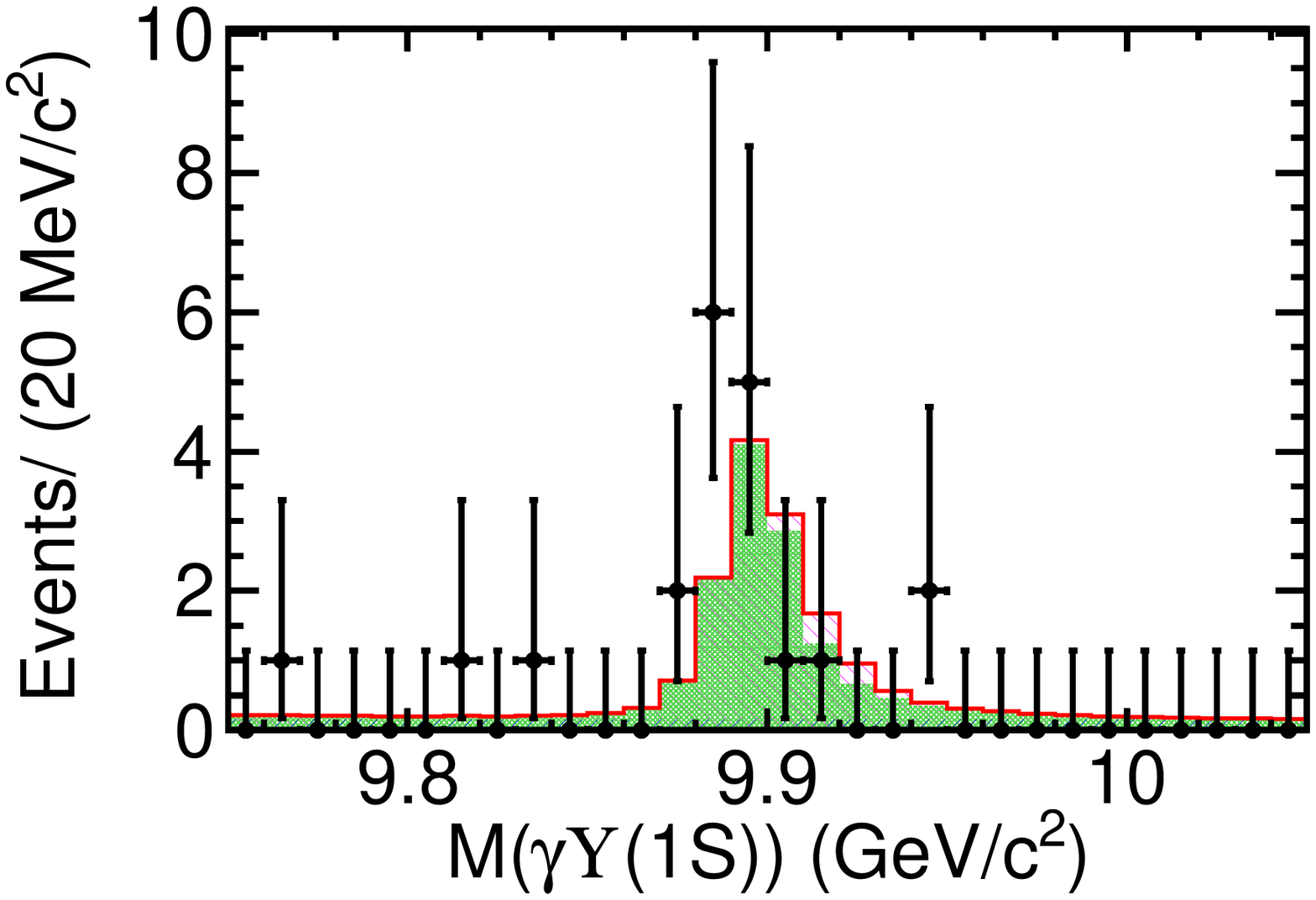}
\caption{{A scatter plot of $M(\pipipi)$ versus
$M(\gamma\Upsilon(1S))$ from data (top left), and the projections
of the 2D fit for events in the $\chi_{bJ}$ signal region (top right), in the $\omega$ signal region (bottom left), and out of the $\omega$ signal region (bottom right).}}\label{fig:2DFit}
\end{figure}

The $\chi_{bJ}$ candidates are reconstructed with the selected
$\Upsilon(1S)$ and the photon not used to form a $\pi^0$ candidate.
The invariant mass of $\pipipi$ ($M(\pipipi)$) versus the corrected invariant mass of $\gamma\Upsilon(1S)$
($M(\gamma\Upsilon(1S))\equiv M(\gamma\LL)-M(\LL)+m_{\Upsilon(1S)}$) is
shown in Fig.~\ref{fig:2DFit} for the sum of the data samples in
the $\yy$ energy region, which is defined as $E_{\rm c.m.} >
10.96$~GeV. Clusters of events for the production of
$\chi_{bJ}$ can be seen both when $M(\pipipi)$ is
in the $\omega$ mass region ([0.75, 0.81]~GeV/$c^2$) and at higher masses ($>0.81~\rm GeV/c^2$). For events
having $M(\pipipi)$ in the $\omega$
mass region, the $\chi_{b2}$ signal is dominant while for signal
events with higher $\pipipi$ masses, the $\chi_{b1}$ signal is
dominant.
The background in this case comes predominantly from false $\pi^0$ candidates produced by combinatorial photons.
%

An unbinned two-dimensional (2D) extended maximum likelihood fit
to the $M(\pipipi)$ and $M(\gamma\Upsilon(1S))$ distributions of the candidate events is applied to
determine the numbers of $\omega\chi_{bJ}$ and $\pipipi\chi_{bJ}$
events. In the fit, the shapes of $\omega\chi_{bJ}$ and
$\pipipi\chi_{bJ}$ obtained from MC simulation are used to
describe the signals, and a 2D function $f(x,y)=ax+by$
($x=M(\gamma\Upsilon(1S))$ and $y=M(\pipipi)$), is used to
fit the background.
Here the $\pipipi\chi_{bJ}$ MC sample is generated following a four-body phase space (PHSP) distribution,
and this process is denoted as $(\pipipi)_{\text{non-}\omega}\chi_{bJ}$.
The projections of the fit results for events in the $\chi_{bJ}$ signal region ($M(\gamma\Upsilon(1S))\in[9.87,9.93]~\rm GeV/c^2$), in the $\omega$ signal region, and in the region above the $\omega$ mass are also shown in Fig.~\ref{fig:2DFit}.
The statistical significances for $(\pipipi)_{\text{non-}\omega}\chi_{b1}$, $(\pipipi)_{\text{non-}\omega}\chi_{b2}$, $\omega\chi_{b1}$ and $\omega\chi_{b2}$ are $5.3\sigma$, $0.0\sigma$, $0.0\sigma$ and $2.5\sigma$, respectively.
The significances are calculated based on the change in likelihood when the signal yield is set to zero in the fit~\cite{significance}.
The signal yields for $(\pipipi)_{\text{non-}\omega}\chi_{b1}$ and $\omega\chi_{b2}$ are $19.6\pm 5.3$ and $7.8\pm 3.2$, respectively, and the signal yields for
$\omega\chi_{b1}$ and $(\pipipi)_{\text{non-}\omega}\chi_{b2}$ are consistent with zero.
Then we assume that either the processes $(\pipipi)_{\text{non-}\omega}\chi_{b1}$ and $(\pipipi)_{\text{non-}\omega}\chi_{b2}$ exist at the same time, or the processes $\omega\chi_{b1}$ and $\omega\chi_{b2}$ exist at the same time, and the fit is repeated.
The statistical significances for $(\pipipi)_{\text{non-}\omega}\chi_{bJ}$ and $\omega\chi_{bJ}$ are $6.1\sigma$ and $4.0\sigma$, respectively.
The changes on the signifiances arise from the similarity in signal shapes between $(\pipipi)_{\text{non-}\omega}\chi_{b1}$ and $(\pipipi)_{\text{non-}\omega}\chi_{b2}$, and between $\omega\chi_{b1}$ and $\omega\chi_{b2}$.
Thus, evidence for $\omega\chi_{bJ}$ has been found, but we cannot determine whether the events are from $\omega\chi_{b1}$ or $\omega\chi_{b2}$.
We also use other forms of background descriptions.
Changes in the signal yields and significances are negligible.

In order to study the energy dependent cross section of $\pipipi\chi_{b1}$ and
$\pipipi\chi_{b2}$ events, we extract the observed signal yields $N_{\rm obs}$ with
data samples listed in Table~\ref{tab:XS}.
Because of the limited statistics for most
energy points, we do not perform a 2D fit as for the summed sample, nor do
we separate $\pipipi$ into $\omega$ and non-$\omega$,
nor $\gamma\Upsilon(1S)$ into $\chi_{b1}$ and $\chi_{b2}$.
The number of $\chi_{bJ}$ signal events in each sample is computed using the formula:
$N_{\rm obs}=N_{\rm sig}-N_{\rm side}$, where $N_{\rm sig}$ is the number of
events in the $\chi_{bJ}$ signal region and $N_{\rm side}$ is that in
the sideband region. Here the signal
region is defined as $M(\gamma\Upsilon(1S))\in$ [9.852, 9.952]~GeV/$c^2$, while
the sideband region is [9.77, 9.82] and [9.98, 10.03]~GeV/$c^2$.

The Born cross sections are calculated with
 \begin{equation}
\sigma^{\rm Born}=\frac{N_{\rm obs}}
{\epsilon\B_{\rm inter}\mathcal{L}(1+\delta)/\left| 1-\Pi\right|^{2}},
 \end{equation}
where $\epsilon$ is the reconstructed efficiency obtained from a series of signal MC samples, $\B_{\rm inter}$ is the corresponding product of intermediate decay branching fractions, $\mathcal{L}$ is the integrated luminosity, $(1+\delta)$ is the ISR correction factor, and ($1/\left| 1-\Pi\right|^{2}$) is the vacuum polarization factor~\cite{VPfactor}.
We use the weighted branching fraction $\B_{\rm weighted}=\B(\chi_{b1}\to\gamma\Upsilon(1S))\cdot f+\B(\chi_{b2}\to\gamma\Upsilon(1S))\cdot(1-f)$,
where $f=N_{1}/(N_{1}+N_{2})=0.74\pm0.06$ is the fraction of $\chi_{b1}$ in the process $\EE\to\pipipi\chi_{bJ}$
near the $\y$ peak~\cite{hadtrans3}.
In order to
estimate the ISR correction factors, we use
\begin{equation}
    1+\delta=\frac{\int^{1-\frac{m_{0}^2}{s}}_{0} G_{\rm BW}(s(1-x))F(x,s) dx}
    {G_{BW}(s)},
\end{equation}
where $m_{0}$ is the mass threshold of $\pipipi\chi_{bJ}$, $F(x,s)$
is the radiative function~\cite{ISRfun} and $G_{\rm BW}(s)$ is the
Breit-Wigner (BW) function,
 \begin{equation}
    G_{\rm BW}(s)=\frac{12\pi\Gamma_{ee}\cdot\mathcal{B}\cdot\Gamma_{\rm tot}}
    {(s-M^2)^{2}+M^2\Gamma_{\rm tot}^2}\times\frac{\Phi(s)}{\Phi(M)},
 \end{equation}
where $M$ is the nominal mass of $\y$ or $\yy$, $\Gamma_{\rm tot}$ is the total width, $\Gamma_{ee}$ is the partial decay width of $\EE$ channel,
$\mathcal{B}$ is the branching fractions of $\pipipi\chi_{bJ}$ channel, and $\Phi$ is given by considering $\pipipi$ as a wide resonance with mass distribution generated in four-body $\EE\to\pipipi\chi_{bJ}$ phase space.

\begin{figure}[htbp]
\centering
    \includegraphics[width=0.50\textwidth]{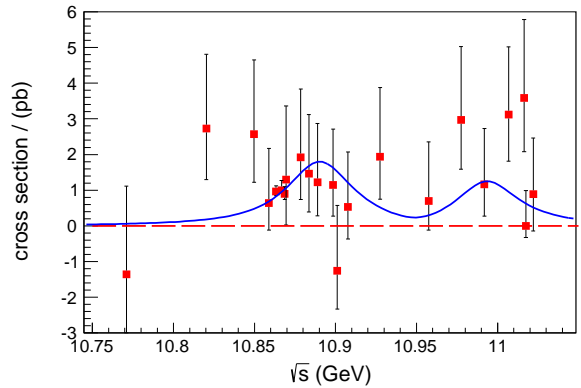}
\caption{Fit to the cross sections of $\EE\to
\pipipi\chi_{bJ}$ as described in the text. The red boxes with error
bars are the cross sections of $\EE\to
\pipipi\chi_{bJ}$ and the solid blue curve
is the fit.}\label{fig:FitXS}
\end{figure}

The energy-dependent cross sections for $\EE\to \pipipi\chi_{bJ}$
are listed in Table~\ref{tab:XS} and plotted
in Fig.~\ref{fig:FitXS}.
A maximum likelihood fit of the cross sections is performed.
The likelihood for the three data samples of larger integrated luminosity around 10.865 GeV is calculated assuming the number of signal events follows the Gaussian distribution:
 \begin{equation}
  L(\mu_{\rm sig};N_{\rm obs},\sigma)=\frac{1}{\sqrt{2\pi}\sigma'}e^{-\frac{(\mu_{\rm sig}-N_{\rm obs})^2}{2\sigma'^2}},
\end{equation}
where $\mu_{\rm sig}$ is the number of expected signal events, and $\sigma'$ is the statistical uncertainty of $N_{\rm obs}$.
For the other samples, the likelihood is calculated assuming the number of signal events
follows the Poisson distribution:
\begin{equation}
\begin{aligned}
    &L(\mu_{\rm sig};N_{\rm sig},N_{\rm side})= \\
    &\int^{\infty}_{0}P(N_{\rm sig};\mu_{\rm sig}+\mu_{\rm bkg})P(N_{\rm side};\mu_{\rm bkg})d\mu_{\rm bkg},
\end{aligned}
\end{equation}
where $P(N;\mu)=\frac{1}{N!}\mu^{N}e^{-\mu}$ is the probability density function of the Poisson distribution, and $\mu_{\rm bkg}$ is the number of expected background events.
The fit function is a coherent sum of
two BW amplitudes in the form of Eq.\ (3) for $\y$ and $\yy$, and the masses
and widths are fixed to their world average
values~\cite{PDG} while the corresponding products $\Gamma_{ee}\cdot\B$ are
left free. The fit results are shown in Fig.~\ref{fig:FitXS}.
Two solutions are found that differ in phase, but
the resulting $\Gamma_{ee}\cdot\mathcal{B}$ are consistent with each other.
The obtained product branching fractions are
$\B(\y\to\EE)\cdot\B(\y\to\pipipi\chi_{bJ})=(15.3\pm3.7)\times10^{-9}$,
$\B(\yy\to\EE)\cdot\B(\yy\to\pipipi\chi_{bJ})=(18.3\pm9.0)\times10^{-9}$,
where the errors are statistical.
We also try to introduce a coherent continuum component into the fit, but the significance of this hypothesis
is only $1.4\sigma$. Thus the continuum component is not included in the nominal fit.
The introduction of the continuum term results in a change of the $\y$ product branching fraction of $12.6\times10^{-9}$ and that of the $\yy$ product branching fraction of $12.8\times10^{-9}$, which are taken as systematic uncertainty due to ``continuum contribution".

%

There are several sources of systematic error in the cross
section measurements, and most of the uncertainties are similar to the
previous work~\cite{hadtrans3}, including tracking
efficiency (1.0\% per pion and kaon track and 0.35\% per lepton),
PID efficiency (1.3\% per pion and 1.6\%
per lepton), photon energy resolution calibration (1.1\%),
$\pi^0$ selection (2.2\%), 5C kinematic fit (4.2\%), and trigger simulation (3.0\%).
The uncertainty from luminosity is 1.5\%~\cite{pipihb}.
Comparing the reconstruction efficiency with the ISR process in
{\sc evtgen} with the efficiency without the ISR process added to {\sc evtgen}, but still corrected for with
the ISR correction factor, yields an uncertainty of 1.0\%. The corresponding
uncertainty from the branching fractions of $\chi_{bJ}\to\gamma\Upsilon(1S),\Upsilon(1S)\to\LL$
is 8.2\%~\cite{PDG}. The total systematic
uncertainty, 11.9\%, is obtained by adding all the above results in quadrature.

The systematic uncertainty in the measured branching fractions rise from the
cross section measurements and the fit to those cross sections.
The systematic uncertainties in the fit to the cross sections mainly come from the
parametrization of the BW function, PHSP factor, resonance parameters,
and the possible continuum contribution. The first is estimated by replacing
the constant width with an energy dependent width $\Gamma_{\rm
tot}=\Gamma^{0}_{\rm tot}\cdot\Phi(\sqrt{s})/\Phi(M)$.
The second source is estimated by replacing the PHSP factor of $\pipipi\chi_{bJ}$ with
the two-body PHSP factor of $\omega\chi_{bJ}$.
The third source is estimated by varying the resonance parameters
$\y$ and $\yy$ within $\pm1\sigma$.
The final systematic uncertainty is estimated by adding a coherent continuum contribution
to the fit function. The changes of
the branching fractions are taken as the symmetrized
systematic uncertainty. The details are listed in
Table~\ref{tab:sysErrY}.

\begin{table}[htbp]
\caption{Summary of the absolute systematic uncertainties in product branching
fractions ($\times10^{-9}$), where $\B(10860,11020)$ represent $\B(\Upsilon(10860,11020)\to\EE)\cdot\B(\Upsilon(10860,11020)\to\pipipi\chi_{bJ})$. } \label{tab:sysErrY}
\begin{tabular}{ c c c  }
\hline
\hline
    $\pipipi\chi_{bJ}$  & $\B(10860)$ & $\B(11020)$ \\
\hline
    Cross sections        &  $1.8$       & $2.1$ \\
    BW parametrization    &  $0.6$       & $0.4$ \\
    PHSP factor           &  $0.6$       & $0.2$ \\
    Resonance parameters  &  $2.4$       & $1.6$ \\
    Continuum contribution&  $12.2$       & $12.6$ \\
\hline
    Sum                   &  $12.6$       & $12.8$    \\
\hline
\end{tabular}
\end{table}

By using $\B(\y\to\EE)=(6.1\pm1.6)\times10^{-6}$ and
$\B(\yy\to\EE)=(2.1^{+1.1}_{-0.6})\times10^{-6}$~\cite{PDG}, we
obtain
$\B(\y\to\pipipi\chi_{bJ})=(2.5\pm0.6\pm2.1\pm0.7)\times10^{-3}$,
$\B(\yy\to\pipipi\chi_{bJ})=(8.7\pm4.3\pm6.1^{+4.5}_{-2.5})\times10^{-3}$,
where the first errors are statistical, the second are
systematic errors combined from the cross section measurements and
line shape fit, and the third result from the branching
fractions of $\y$ and $\yy\to \EE$~\cite{PDG}.


To reconstruct $\EE\to \phi\chi_{bJ}$, we require at least two
kaons in one event. There is no requirement on the number
of photons, but a list of photon candidates is created in one event satisfying
$|M(\gamma\gamma_{2})-m_{\pi^0}|>13$~MeV/$c^2$, where $\gamma_2$
is any other photon in the event with $E_{\gamma_2}>0.1$~GeV, and $m_{\pi^0}$ is the
nominal mass of the $\pi^0$.
The data are divided into two categories.
One includes events when one of the photons in the above
list satisfies $M_{\rm recoil}(\gamma\KK)\in [9.42,~9.50]~\rm
GeV/c^2$, i.e., in the $\Upsilon(1S)$ mass region, to tag $\chi_{bJ}\to\gamma\Upsilon(1S)$ events;
the other includes all other events, to tag $\chi_{bJ}\to$non-$\gamma\Upsilon(1S)$ events.
Here $M_{\rm recoil}(\gamma\KK)$ is the recoiling mass of the $\gamma\KK$ system in the $\EE$ c.m. frame.

We use the figure of merit, $\mathcal{S}/\sqrt{\mathcal{S}+\mathcal{B}}$, to
optimize the $\KK$ invariant mass window requirement.
Here $\mathcal{S}$ is the
reconstructed number of signal events obtained from MC
simulation of the signal process, $\yy\to\phi\chi_{bJ}$ with $\phi\to\KK$, $\chi_{bJ}\to$~anything, in the signal region, [9.88, 9.93]~GeV/$c^2$.
The number is normalized according to the theoretical calculation of the branching fraction of $\yy\to\phi\chi_{bJ}$~\cite{Theory} and the total $\yy$ events in our data sample.
$\mathcal{B}$ is the number of
background events in the signal region in the generic MC sample with the c.m.\ energy shifted to $11.022$~GeV.
For the first category, we require $M(\KK)$ to be within $m_{\phi}\pm7.5~\rm MeV/c^2$, and for category two, we require
$M(\KK)$ to be within $m_{\phi}\pm 7.0$~MeV/$c^2$, where
$m_{\phi}$ is the nominal mass of $\phi$~\cite{PDG}. The $\phi$
mass sideband region is defined as $M(\KK) \in [1.000,~1.005]$ or
$[1.035,~1.040]$~GeV/$c^2$.
There is no evidence for the $\chi_{bJ}$ signal in the $\phi$ mass sideband events, nor
in the generic MC sample (significance is less than $0.1\sigma$ from the fit) mentioned above.

\begin{figure}[htbp]
\centering
    \includegraphics[width=0.5\textwidth]{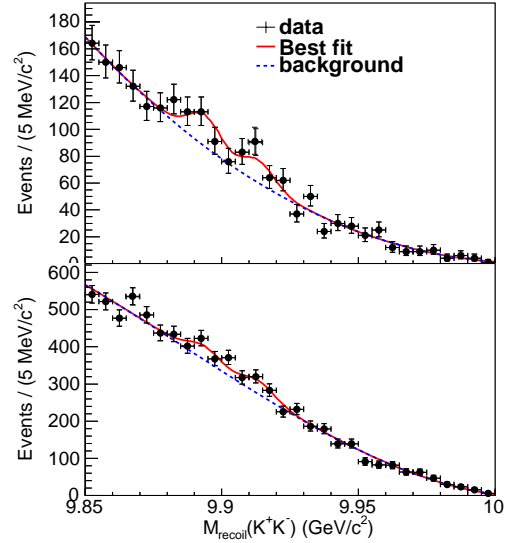}
\caption{The simultaneous fit results for data having $M_{\rm recoil}(\gamma\KK)$, with the recoiling mass of $\gamma\KK$, in the
$\Upsilon(1S)$ mass window (up) and out of the $\Upsilon(1S)$ mass window (down). Dots with error
bars are data, the red solid lines are the best fit, and blue
dashed lines are backgrounds.}\label{fig:SimuFit}
\end{figure}

After applying all the selection criteria, the recoil mass spectra
of $\phi$ as a function of the initial beam four-momenta from both data categories are shown in
Fig.~\ref{fig:SimuFit} for the sum of data in the energy region $\sqrt{s}=10.96$---$11.05$~GeV.
We perform a simultaneous unbinned maximum likelihood fit to the $\phi$ recoil mass spectra
with the signal shapes from the simulated signal MC shapes, and a background
shape obtained from data with the following procedure:
a series of shapes are obtained from $\Upsilon(5S)$ data, where, in
calculating the $\KK$ recoil mass, the c.m. energy is
changed to that of each individual data point, and summing up the shapes according to the luminosity.
The ratios of the numbers
of $\chi_{b1}$ or $\chi_{b2}$ events in the two categories are fixed according to the
expected branching fractions of $\chi_{b1}$ or $\chi_{b2}\to\gamma\Upsilon(1S)$~\cite{PDG} and the efficiencies. The
fit results, which yield $\chi^2/\rm ndf=104.2/55=1.9$,
are shown in Fig.~\ref{fig:SimuFit}.
According to the fit, $(1.5\pm 0.5)\times10^3$~$\chi_{b1}$ and $(2.4\pm
0.5)\times10^3$~$\chi_{b2}$ events are produced.
The statistical significances are found to be $3.3\sigma$ and
$4.8\sigma$ for $\chi_{b1}$ and $\chi_{b2}$, respectively.

When we vary the background shape by multiplying the nominal background
shape with a first, second-, or third-order polynomial, the
smallest significances of the $\chi_{b1}$ and $\chi_{b2}$ signals
are found to be $2.6\sigma$ and $2.1\sigma$, respectively (multiplying
by the third-order polynomial), yielding $\chi^2/\rm ndf=43.6/49=0.89$.
The most conservative upper limits on the numbers of
produced signal events in all the above tests are reported. After
considering the systematic uncertainty which we discuss
later, the upper limits for the produced numbers of
$\phi\chi_{b1}$ and $\phi\chi_{b2}$ signal events are determined
to be $2.2\times10^3$ and $3.1\times10^3$ at 90\% confidence level
(C.L.), respectively.
The upper limits on the Born
cross sections of $\EE\to \phi\chi_{b1}$ and $\phi\chi_{b2}$
are $0.7$ and $1.0~\rm pb$, respectively, averaged over the $\yy$ region, specifically
$\sqrt{s}=10.96$---$11.05$~GeV.
The calculation is based on Eq.~(1), where the reconstruction efficiency, ISR correction factor, and vacuum polarization factor
are averaged with weights according to the luminosity of each sample.

The sources of systematic uncertainties in the $\phi\chi_{bJ}$ cross
section measurement are similar to those of the $\pipipi\chi_{bJ}$ modes,
including the tracking efficiency, PID, photon detection,
luminosity, trigger simulation, ISR correction, $\phi$ mass
window, and intermediate branching fraction. Most of these have
been discussed in the $\pipipi\chi_{bJ}$ analysis. The
uncertainty from the $\phi$ mass window requirement is found to
be negligible by studying the consistency of the $\KK$ invariant
mass between data and MC simulation. The uncertainty from the branching fraction
of $\phi\to\KK$ is 1.0\%~\cite{PDG}.
The total systematic
uncertainty for the cross section measurement is thus, combining all uncertainties in quadrature,
5.5\% for $\EE\to\phi\chi_{b1}$ or $\phi\chi_{b2}$.

In summary, using the energy scan data in the vicinity of
the $\yy$ resonance, we observe the $\EE\to (\pipipi)_{\rm non-\omega}\chi_{b1}$ process with significance
of $5.3\sigma$.
Evidence for $\omega\chi_{bJ}$ is found but we cannot tell whether it is $\omega\chi_{b1}$ or $\omega\chi_{b2}$.
The limited statistics prevents us from drawing a conclusion
concerning the origin of the signal events, that is, whether they arise
from bottomonium decay, continuum production,
or both.
Since no continuum production of a multibody final state with a bottomonium is known,
it is natural to assume that the origin of the signal is bottomonium.
Under this assumption, the branching fractions are $\B(\y\to\pipipi\chi_{bJ})=(2.5\pm0.6\pm2.1\pm0.7)\times10^{-3}$, which is compatible with the previous measurement~\cite{hadtrans3}, and $\B(\yy\to\pipipi\chi_{bJ})=(8.7\pm4.3\pm6.1^{+4.5}_{-2.5})\times10^{-3}$, which is compatible with the theoretical predictions~\cite{Theory}.
Based on the 2D fit with summed data, the relative magnitude $R_{21}(\omega)\equiv\frac{\B(\yy\to\omega\chi_{b2})}{\B(\yy\to\omega\chi_{b1})}$ can be estimated to be $0.4\pm0.2$, where the common systematic uncertainties cancel.

The processes $\EE\to \phi\chi_{bJ}$ are also searched for in data within
$\sqrt{s}=10.96$---$11.05$~GeV, with no significant signals being observed.
We report upper limits on the Born cross sections of $\EE\to\phi\chi_{b1}$
and $\phi\chi_{b2}$ as $0.7$ and $1.0~\rm pb$ at
90\% C.L., respectively. Compared with the total cross section of
$\EE\to \yy$, these upper limits correspond to $\yy$ decay
branching fractions of order $10^{-3}$, well above the theoretical
predictions of order $10^{-6}$~\cite{Theory}.

Our measurement of the transition rate agrees with the expectation of Ref.~\cite{Theory},
but the measured relative magnitudes $R_{21}(\omega)$ are significantly less than the theoretical predictions, which should be more reliable than the branching fraction predictions.
This may inspire theorists to further investigate the discrepancy between the experimental measurement and the theoretical calculation


We thank Dr. K. Li for his fantastic idea and kind help.
We thank the KEKB group for excellent operation of the
accelerator; the KEK cryogenics group for efficient solenoid
operations; and the KEK computer group, the NII, and
PNNL/EMSL for valuable computing and SINET5 network support.
We acknowledge support from MEXT, JSPS and Nagoya's TLPRC (Japan);
ARC (Australia); FWF (Austria); NSFC and CCEPP (China);
MSMT (Czechia); CZF, DFG, EXC153, and VS (Germany);
DST (India); INFN (Italy);
MOE, MSIP, NRF, RSRI, FLRFAS project and GSDC of KISTI (Korea);
MNiSW and NCN (Poland); MSHE and RSF, Grant 18-12-00226 (Russia);
ARRS (Slovenia);
IKERBASQUE and MINECO (Spain);
SNSF (Switzerland); MOE and MOST (Taiwan); and DOE and NSF (USA).

\end{document}